\documentclass[10pt]{llncs}
\usepackage{amsmath}
\usepackage{amsfonts}
\usepackage{amssymb}
\usepackage{algorithmic}
\usepackage{psfrag}
\usepackage{subfig}
\usepackage{wrapfig}
\usepackage{url}
\usepackage{cite}

  \usepackage[pdftex]{epsfig}
  \usepackage[pdftex]{hyperref}

\newcommand{\Z}{\mathbb{Z}}

\newcommand{\pfunc}[3]{#1 : #2 \dashrightarrow #3 }

\begin{document}

\title{Simulation of Self-Assembly in the Abstract Tile Assembly Model with ISU TAS\thanks{This research was supported in part by National
   Science Foundation Grants 0652569 and 0728806}}%
\author{Matthew J. Patitz}
\institute{Department of Computer Science \\
Iowa State University \\
Ames, IA 50011, U.S.A. \\%
\email{mpatitz@cs.iastate.edu}%
}

\date{}

\clearpage

\maketitle
\begin{abstract}
Since its introduction by Erik Winfree in 1998, the abstract Tile Assembly
Model (aTAM) has inspired a wealth of research.  As an abstract model for
tile based self-assembly, it has proven to be remarkably powerful and
expressive in terms of the structures which can self-assemble within it.
As research has progressed in the aTAM, the self-assembling structures being
studied have become progressively more complex.  This increasing complexity,
along with a need for standardization of definitions and tools among researchers,
motivated the development of the Iowa State University Tile Assembly Simulator
(ISU TAS).  ISU TAS is a graphical simulator and tile set editor for
designing and building 2-D and 3-D aTAM tile assembly systems and simulating
their self-assembly.  This paper reviews the features and functionality of ISU
TAS and describes how it can be used to further research into the complexities
of the aTAM.  Software and source code are available at \url{http://www.cs.iastate.edu/~lnsa}.
\end{abstract}

\clearpage

\section{Introduction}

Numerous advances in science and technology are fueling rapid
transformations in the types of materials and structures that
can be manufactured and the processes for manufacturing them.
On one hand, such advances allow us to build ever larger structures
due to the stronger, lighter, and more durable materials that can
be fabricated.  On the other hand, breakthrough techniques for
controlling matter at microscopic levels have led to the amazing
trend of shrinking products, such as the incredible reduction in
the size of computer processors as dictated by Moore's Law.

While there are numerous manufacturing techniques for building this
diversity of products, they tend to have fundamental similarities.
Namely, they begin with relatively large chunks of matter which
are then cut, molded, or otherwise shaped in a manner controlled
by high-level, outside manipulation.  Until now, this `top down'
approach to manufacturing has been successful and sufficient, but
now, in order to meet ever more demanding challenges, research is
being conducted on an entirely new paradigm, a `bottom up' approach.

Self-assembly is the process by which units of matter autonomously
combine to form structures.  By pursuing methods of nanoscale
self-assembly, researchers are hoping to eventually create systems
where they design molecular pieces, mix them together, and watch as
the desired structures spontaneously arise from the pieces.  Such
systems have the potential to generate materials and both nanoscale
and macro-scale structures with desired qualities far surpassing
those achievable today.

In the early 1980's, Ned Seeman began pioneering reserach into
self-assembly systems based on DNA molecules \cite{Seem82}.  Later,
in 1998, Erik Winfree created an abstraction of such systems called
the Tile Assembly Model \cite{Winf98}.  The Tile Assembly Model (TAM),
an effectivization of Wang tiling \cite{Wang61},
is based on DNA molecules which are structured so that they behave
like square tiles which have glues on their edges that allow them
to stick to tiles which have matching glues.  Such tile assemblies
have actually been created in vitro, such as in \cite{RoPaWi04}.

Under the broad umbrella of the TAM, there are kinetic versions (kTAM)
which deal with molecular concentrations, reaction rates, and other
physical variables, and there are abstract versions (aTAM) which further
abstract the physical characteristics of the model.  In this paper we
will focus on versions of the aTAM, which we will briefly outline in
Section \ref{TAM}.

Assembled structures, or assemblies, in the aTAM start from predefined
`seed' structures and can grow to produce finite or infinite structures
which represent predefined shapes or represent the outputs of computations.
This model has proven to be remarkably powerful and expressive in terms
of the types of assemblies which can be produced and the computations
which can be performed, and much research has been done to explore this
potential.

As this research into the aTAM has progressed, the tile sets being considered
and the assemblies studied have steadily increased in size and complexity.
Additionally, various alternatives to the original aTAM have been proposed
and investigated.  This growth has resulted in the need for powerful
software tools that can aid in the development and visualization of
such sophisticated systems, as well as standardizing results for valid
discussion and comparison across diverse approaches.  For just such reasons,
the Iowa State University Tile Assembly Simulator, or ISU TAS, was
developed.

ISU TAS is a freely available, open source, cross-platform software package
which provides a full development and simulation environment for the abstract
Tile Assembly Model.  It provides the ability to create and edit both 2-D and
3-D tile assembly systems in a graphical framework, and then simulate the growth
of the ensuing assemblies.  Various parameters of the particular version of
the aTAM being used can be specified, and several features are provided to
help in debugging tile assembly systems.

In this paper, after sketching a definition of the aTAM, we briefly discuss
a prior simulator, Xgrow, and its
abilities and limitations.  We then give a more detailed breakdown of ISU TAS
and its current features and functionality, with sections focusing on
both the tile set editor and the simulator.  Next, we briefly mention some
tools we also make available for algorithmically generating tile assembly
systems which can be simulated by ISU TAS.  Finally, we mention future
direction and additional features and functionality that we hope to implement
in ISU TAS.

\section{The Abstract Tile Assembly Model}\label{TAM}

This section provides a very brief overview of the most commonly
used variation of the aTAM, which is supported by ISU TAS. Where
deviations which are also supported by ISU TAS are discussed in
the proceeding sections, they will be briefly defined. See
\cite{Winf98,RotWin00,Roth01,jSSADST} for other developments of the
model. Our notation is that of \cite{jSSADST}.

We work in the discrete space $\Z^n$, where $n = 2$ or $3$.   We write $U_n$
for the set of all {\it unit vectors}, i.e., vectors of length 1 in
$\mathbb{Z}^n$.  Intuitively, a tile type $t$ is a unit square (or cube if
$n = 3$) that can be translated, but not rotated, having a well-defined ``side $\vec{u}$''
for each $\vec{u} \in U_n$. Each side $\vec{u}$ of $t$ has a ``glue'' of ``color''
$\textmd{col}_t(\vec{u})$ - a string over some fixed alphabet $\Sigma$ - and
``strength'' $\textmd{str}_t(\vec{u})$ - a natural number - specified by its
type $t$. Two tiles $t$ and $t'$ that are placed at the points $\vec{a}$ and
$\vec{a}+\vec{u}$ respectively, {\it bind} with {\it strength}
$\textmd{str}_t\left(\vec{u}\right)$ if and only if
$\left(\textmd{col}_t\left(\vec{u}\right),\textmd{str}_t\left(\vec{u}\right)\right)
=
\left(\textmd{col}_{t'}\left(-\vec{u}\right),\textmd{str}_{t'}\left(-\vec{u}\right)\right)$.

A {\it tile assembly system} ({\it
TAS}) is an ordered triple $\mathcal{T} = (T, \sigma, \tau)$, where
$T$ is a finite set of tile types, $\sigma$ is a seed assembly with
finite domain, and $\tau \in \mathbb{N}$ is the temperature.  
An {\it assembly} is a partial function $\pfunc{\alpha}{\Z^n}{T}$,
and an {\it assembly sequence} is a (possibly
infinite) sequence $\vec{\alpha} = ( \alpha_i \mid 0 \leq i < k )$
of assemblies in which $\alpha_0 = \sigma$ and each $\alpha_{i+1}$
is obtained from $\alpha_i$ by the ``$\tau$-stable'' addition of a
single tile.  An assembly is $\tau$-{\it stable} if it cannot
be broken up into smaller assemblies without breaking bonds whose strengths
sum to at least $\tau$.

Self-assembly begins with a {\it seed assembly} $\sigma$ and
proceeds asynchronously and nondeterministically, with tiles
adsorbing one at a time to the existing assembly in any manner that
preserves stability at all times. The {\it $\tau$-frontier}, or simply {\it frontier},
of an assembly $\alpha$ is the set of all positions at which a tile from $T$
can be ``$\tau$-stably added'' to the assembly.  (Note that in ISU TAS,
the definition of the term {\it frontier} is relaxed to refer to locations
which have no tiles but have tiles in adjacent locations such that the sum
of their incident glue strengths is at least $\tau$, although there may or
may not be a tile type in $T$ which can validly bind.) An assembly is
called {\it terminal} when its frontier is empty, meaning that no more tiles
can bind.

Soloveichik and Winfree defined the notion of {\it local determinism},
which is a powerful tool for proving the correctness of tile assembly
systems, and we refer the reader to \cite{SolWin07} for a full definition.

%
\section{Previous Work}

The Xgrow Simulator \cite{Xgrow} is a graphical simulator for both
the aTAM and the kTAM.  It was written by the DNA and Natural
Algorithms Group, headed by Erik Winfree, at the California Institute
of Technology.  It is available for download, along with source code,
at the following URL: \url{http://www.dna.caltech.edu/Xgrow}.

Xgrow is written in C for X Windows environment and supports
a wide range of options for controlling various parameters of
the tile assembly systems it simulates.  It also allows for
modification of the environment and the assembly dynamically,
while an assembly is growing.  This functionality allows researchers
to better understand the interplay of the many factors that
influence assembly in the aTAM and kTAM.

However, while Xgrow is very useful for gaining high-level insights
into how tile-based self-assembly works, it doesn't provide the
ability to inspect assemblies at the level of individual tiles, or
provide tools for editing or debugging tile assembly systems.  This
makes it difficult to design complex new tile assembly systems, and
was the main motivation for the creation of ISU TAS.

\section{ISU TAS Overview}

The Iowa State University Tile Assembly Simulator
is an integrated platform for designing, simulating, testing,
and debugging tile assembly systems in the abstract Tile
Assembly Model.  The software and source code are available
for download from \url{http://www.cs.iastate.edu/~lnsa}.

ISU TAS is broken into two main components, the
simulator and the tile set editor, which will each be covered
in detail in proceeding sections.  In this section we provide
an overview of the underlying architecture as well as
describing a subset of its features.

\subsection{Code Base}

ISU TAS is written in C++, and the source code is open
source and freely available.  It is built upon the wxWidgets
toolkit \cite{wxWidgets}, which is a set of cross-platform
C++ libraries, and can be built and run on both Windows and
Linux operating systems.  Build scripts are included for
both platforms, and a compiled Windows executable is
available for download.

Additional third party libraries are also utilized to
provide several enhanced features.  To provide 3-D
rendering, ISU TAS makes use of the OpenGL \cite{OpenGL}
and Texfont \cite{Texfont} libraries. RandomLib \cite{RandomLib}
is used for random number generation in order to provide
a uniform distribution.

\subsection{Architecture}

In the aTAM, tile assembly systems are defined by three
parameters:  a tile set, a seed assembly, and a temperature
value.  To facilitate the building of tile assembly systems,
ISU TAS provides tools to graphically build and edit tile
sets, design seed assemblies, and set the temperature parameter.
This functionality is split between two largely disjoint
components, the simulator and the tile set editor.  The simulator
maintains the definition of a full tile assembly system, while
the editor maintains the definition of a separate copy of
the tile set, allowing it to be edited without invalidating
any assembly currently contained within the simulator.  Therefore,
whenever an edited version of the tile set is to be used
within a simulation, the existing assembly in the simulator
is reset to the seed configuration and the tile set from the
editor is copied over the tile set maintained by the simulator.

The simulator and tile set editor each have their own top level
window.  Each of those have dockable sub-windows which can be toggled
on and off to provide additional information.

Tile assemblies and tile sets can be saved to and loaded from
files.  The format is a very simple text file format which
also allows for modification within standard text editors
or easy programmatic generation.

\subsection{Supported Variations of the aTAM}

ISU TAS supports several variations of the aTAM (including a few
which differ from the definition in Section \ref{TAM}).  Some of
the configurable parameters for tile assembly systems include the
following:

\begin{enumerate}
    \item The temperature value can set to any desired positive integer.
    \item At each time step of the simulation, either one frontier location
    can be selected at random into which a fitting tile type is placed, or
    every frontier which exists at the beginning of that time step can be
    filled in a single time step.
    \item Tiles and assemblies can be either 2-dimensional or 3-dimensional.
    \item Values can be specified for the relative concentrations of tile
    types so that, given frontier locations where multiple tile types can
    fit, a particular tile type is selected with probability proportional
    to its concentration value relative to all other fitting tile types.
\end{enumerate}

\section{The Tile Set Editor}

\begin{figure}[htp]
    \begin{center}
        \includegraphics[width=4.8in]{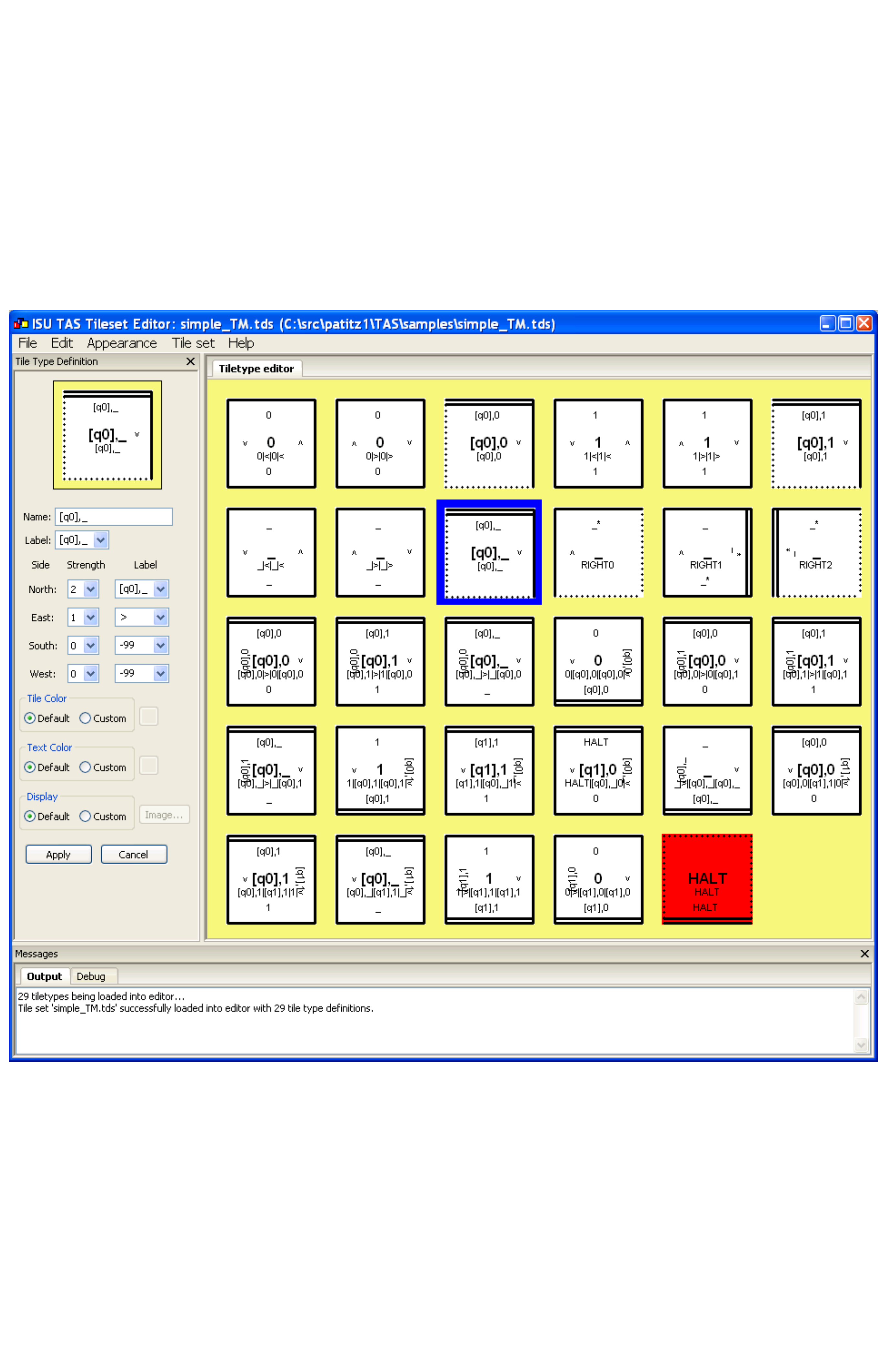}
        \caption{\small \label{fig:editor_win} Tile set editor window}
    \end{center}
\end{figure}

In the tile set editor window, a new tile set can be created
or an existing one loaded.  Figure \ref{fig:editor_win}
shows a screenshot of the tile set editor window. Each tile
type is graphically depicted in the `Tiletype editor'
window, where they can be cut, copied, pasted, and dragged
into different orderings, singly or in selected groups.
If a particular tile type is selected by left-clicking on it,
it is loaded into the `Tile Type Definition' window, where
every attribute of it can be edited.

Additional features of the editor include the ability to:
\begin{enumerate}
    \item Rotate tile types
    \item Search for tile types with attributes matching
    user-specified strings
    \item Search for tile types which can bind to a particular
    side of a selected tile type
    \item Highlight all tiles which are being used (or unused)
    in the current assembly contained in simulator
\end{enumerate}

Additionally the editor automatically highlights any tile
types which are functionally equivalent (they have all of
the same glues).

All modifications to the tile set in the editor are independent
of the tile set loaded into the simulator until the editor's
tile set is manually copied into the simulator, which can be done
via the `Tile set' menu or a toolbar button in the simulator window.
\section{The Simulator}

The simulation window of ISU TAS allows a user to create,
load, and save tile assembly systems, either as a unit or
as separate components.  Simulation can be done one step
at a time or in a fast-forward mode.  Simulation steps
are cached, so they can also be run in reverse.  The
simulation engine is optimized to maximize the speed of
assembly while handling very large tile sets (testing
has been done with tile sets containing over $10,000$
unique tile types).  To provide for maximum simulation
speed, the simulator can be configured to redraw the
display of the assembly only at user-specified intervals.

Seed assemblies can be created by moving the mouse cursor
over the desired coordinates for a seed assembly tile, then
right-clicking and selecting the appropriate tile type from
the menu which appears.

\begin{figure}[htp]
    \begin{center}
        \includegraphics[width=4.8in]{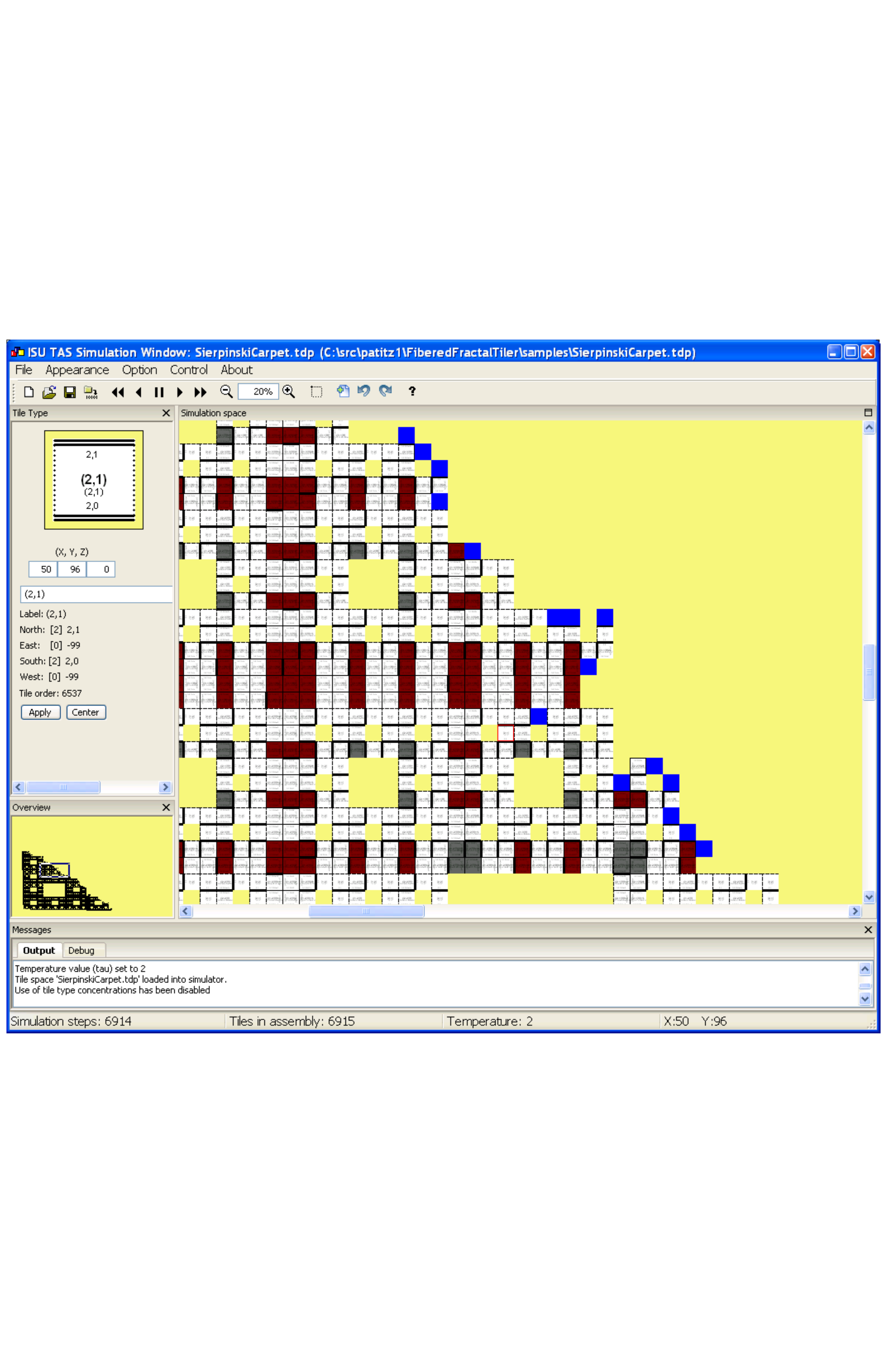}
        \caption{\small \label{fig:sim_win} Simulation window}
    \end{center}
\end{figure}

\subsection{Viewing the Assembly}
Figure \ref{fig:sim_win} shows the simulation window during
the simulation of a two-dimensional tile assembly.  The
`Simulation space' window shows a portion of the current
assembly and allows for arbitrary zoom factors and panning
across the entire assembly for high level viewing of the
assembly or for tile by tile inspection.  It shows
the tiles in the assembly as well as highlighting frontier
locations with blue squares.  The `Overview' window (seen near
the bottom left of Figure \ref{fig:sim_win}) shows a small
version of the entire assembly with a box drawn around the
area that is currently viewable in the `Simulation space'
window.  Clicking within the `Overview' window causes the
`Simulation space' window to automatically scroll so that the
clicked location of the assembly is centered.

When the mouse cursor is moved over the assembly, the `Tile
type' window shows the contents of the location currently
under the cursor, allowing tile types to be clearly seen
without requiring large zoom factors.  Besides the attributes
of the tile type, the time step in which a tile was added
to that particular location is also displayed, along with
its coordinates.

\subsection{Debugging Features}

Due to the immense complexity of many tile assembly systems
being researched, there is a great need for extensive debugging
features that can be used during their development.  Some
of those provided by ISU TAS are listed below.

\begin{enumerate}
    \item Breakpoints can be set to stop fast-forward simulation
    based on any of the three user-specified criteria:
        \begin{enumerate}
            \item A specified number of simulation steps have occurred
            \item A tile is placed in a specified location
            \item A tile of a specified type is placed in any location
        \end{enumerate}
    \item The seed used for random number generation can be retrieved,
    manually set, or automatically generated, and every time an assembly
    is reset it is possible to specify what seed is used in order
    to provide reproducible results when debugging issues that arise
    due to the nondeterminism inherent in the TAM.
    \item Locations which have incident glue strengths equal to or
    greater than $\tau$ are drawn as blue squares since they are
    eligible frontier locations, and any such locations at which
    a tile addition has been attempted but no possible fitting
    tile type was found are drawn as red squares and referred to as
    `dead' frontier locations.
    \item A box-drawing tool can be used which allows frontier
    locations to be selected and toggled `on' or `off' in order to
    restrict assembly growth to particular locations.
    \item An option can be turned on which causes the simulator to
    report every instance in which a tile was added to a location
    in which more than one tile type could have validly been placed
    (a type of non-determinism).
    \item An option can be turned on which causes the simulator to
    report every instance in which a tile was added to a location
    in which it bound with strength greater than $\tau$, which is
    a violation of the first condition of local determinism (Soloveichik
    and Winfree \cite{SolWin07}).
\end{enumerate}

\subsection{3-D Simulation}

\begin{figure}[htp]
    \begin{center}
        \includegraphics[width=4.8in]{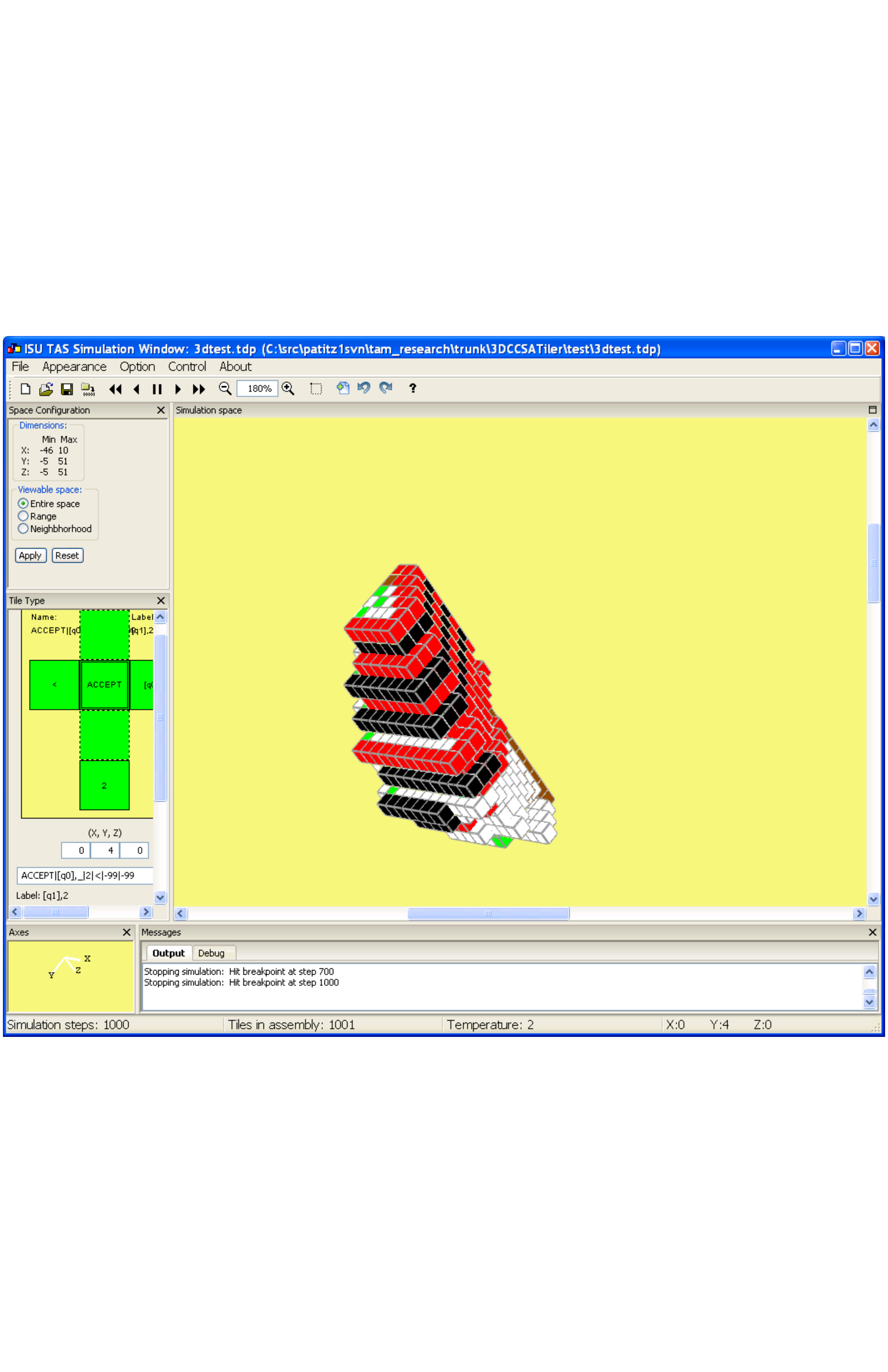}
        \caption{\small \label{fig:sim_3d_win} Simulation window (3-D mode)}
    \end{center}
\end{figure}

Figure \ref{fig:sim_3d_win} shows the simulation window when
the simulator is in 3-D mode.  The differences in 3-D simulation
from 2-D mostly concern the visualization.  In 3-D mode, the mouse
is used to rotate the assembly and view it from different angles.
One additional window, the `Axes' window, displays the current
orientation of the three positive axes, while another, the `Space
Configuration' window, allows user-defined regions (or `slices') of
the assembly to be the only visible portions.  This allows for inspection
of arbitrary pieces of the assembly, even if they are interior and
would otherwise be blocked from view by other portions of the assembly.
Frontier locations are displayed as semi-transparent cubes.  Finally,
the `Tile Type' window now shows an `unwrapped' three-dimensional tile
so that all sides can be simultaneously viewed.

\section{Additional Tools}

In addition to ISU TAS, we also make several programs for
algorithmically generating tile assembly systems used by
ISU TAS, and their source code, available.  They are written
in standard C++ with no requirements on third party libraries
and include a basic library, TileLib, which can be used to
easily create new applications.  Tools for generating tile
sets which act as counters, Turing machine simulations, and
other constructions related to several of our results can
all be downloaded.
\section{Future Work}

While ISU TAS has come a long way toward becoming
a solid and robust environment for designing and
testing tile assembly systems, there remain many
features on our list to implement in the future.

Many more optimizations for 3-D simulation are
required to support large assemblies.  Partial
support for temperature programming has been implemented,
but an efficient way to calculate the fragments of
assemblies which should `melt' off at temperature
increases is needed.  Building and testing of ISU
TAS needs to be performed on Mac OSX in order to
add that as a supported platform.  We also wish to
add support for versions of the kTAM.

With increased interest in, and usage of, ISU TAS we
hope to receive useful feedback and testing that can
enable us continue to provide tools that help further
research in tile-based self-assembly and aid in moving
this theoretical research closer to a physical reality.

\subsubsection*{Acknowledgments}
We thank Scott Summers and Dave Doty for valuable testing, feature requests, and
feedback, as well as tremendous patience during prolonged development and
debugging cycles.

\bibliographystyle{amsplain}
\bibliography{main,dim,random,dimrelated,rbm,tam,code}

\end{document}